# Catalog of Luminous Supersoft X-ray Sources

J. Greiner

Max-Planck-Institut für Extraterrestrische Physik, 85740 Garching, Germany

## 1 Introduction

This catalog comprises an up-to-date list of luminous ($>10^{36}$ erg/s) supersoft X-ray sources. Since most of the new sources are X-ray discoveries, the final inclusion in the group of luminous close binary supersoft sources has to await the optical identification. Only then a distinction is possible among the various and quite different types of objects which show a supersoft X-ray spectrum (i.e. emission only below 0.5 keV): single (i.e. non-interacting) white dwarfs, central stars of planetary nebulae, PG 1159 stars, symbiotic variables, magnetic cataclysmic variables, active galactic nuclei. An exception are F and G stars which also have supersoft X-ray spectra but which can readily recognised due to their optical brightness (low $L_x/L_{opt}$ ratio). Due to this fact of necessary follow-up optical observations, it can well happen that a source is included in the catalog but later turns out to be of a different type. A rather recent example is RX J0122.9–7521 which has long been thought to be a SMC supersoft source (Kahabka et al. 1994), but has been identified as a galactic PG 1159 star (Cowley et al. 1995, Werner et al. 1996).

We include in this catalog accreting binary sources of high luminosity which are thought to be in a state of (steady or recurrent) hydrogen burning. Since CAL 83, the prototype, is known to have an ionisation nebula (Pakull and Motch 1989), and further supersoft binaries are expected to also have one, we include also sources associated with very luminous planetary nebulae. Not included are PG 1159 stars which reach similar magnitudes but form a rather distinct class of presently 27 members (Dreizler et al. 1995). Excluded are also supersoft active galactic nuclei which reach luminosities up to $10^{45}$ erg/s.

The first two supersoft X-ray sources, CAL 83 and CAL 87 (Long et al. 1981), have been discovered with Einstein satellite observations. However, ROSAT observations established these sources as a distinct class, and the majority of the X-ray measurements have been performed with the ROSAT position sensitive proportional counter (PSPC). The PSPC with its spectral resolution of about 50% below 1 keV has been used in nearly all cases to discover the supersoft X-ray spectrum. During the last years the high-resolution imager (HRI) has been used to improve the coordinates of the newly detected sources and to monitor their X-ray intensity. At these soft energies, the HRI countrates are typically a factor of 7.5–8 smaller than those of the PSPC (David et al. 1994, Greiner et al. 1996a).

## 2  Organisation

The catalog is organised as follows:

- The objects appear in order of increasing right-ascension (for equinox 2000.0) though this mixes SMC and M31 sources. Tab. 1 lists all sources according to their host galaxy with some important parameters for quick comparison.
- The original name of the source is written at the top left, and widely known secondary names on the right.
- The first two lines for each source contain the coordinates, the references for these coordinates and the discovery (in case these are different) and the type classification if the source is optically identified. If a source is not optically identified, we mark it as $\boxed{\text{unidentified}}$ below the type identification, and the R.A./Dec. numbers are the X-ray positions with uncertainties as given in the above section of ROSAT related issues. Otherwise, the optical positions are given with typical uncertainties of $1''$. All coordinates are equinox 2000.0. The reference in the "Coordinates" field refers to the most accurate position, i.e. for identified objects the optical position overrides the X-ray position. The "Discovery" reference refers to the first paper which realises the luminous supersoft X-ray emission. Some of the sources have been known already for decades at this time, so the "Coordinates" reference can be much earlier than the "Discovery" reference.
- The next block contains general data which are not specific to any wavelength range. For sources supposed to belong to an external galaxy, the galaxy name is given instead of the presently known distance. This is motivated by recent changes in the distance determination of the LMC which reduced from the standard 55 kpc to 47.3±0.8 kpc (Gould 1995). This in turn also affects the distance to M31 because the distance ratio of LMC and M31 is more accurately known than the corresponding absolute distances. The most recent SMC distance is 57.5 kpc (van den Bergh 1989). The galactic absorbing column ($N_H^{gal}$, Dickey and Lockman 1990) is given for comparison with the values derived from the X-ray fits given in the next block.
- Two blocks follow for data derived from X-ray and optical/UV measurements. Except the brightness estimates, all numbers are referenced. The temperature derived from the X-ray spectra is given as a range including the error estimates and specifying the model used: bb for the blackbody model and wd for white dwarf atmosphere models. It should be noted that for some sources the values for the best-fit absorption are for the blackbody model while the temperature is from the white dwarf model, i.e. this is not consistent (unfortunately the fitted absorbing column in white dwarf models is only rarely given in the literature).
- Finally, the references are given in full with all co-authors, and stating the important pieces of new information (this reflects purely the subjective view of the author, and you should contact me if something is missing or wrong.) The references are sorted in time of appearance, so that the numbers in the data blocks will not change when the catalog is updated.

**Table 1.** Summary of all known supersoft X-ray sources with luminosities above $10^{36}$ erg/s excluding PG 1159-type stars and supersoft AGN. Given are for each source the name (column 1), the best fit X-ray temperature with bb indicating blackbody and wd white dwarf atmosphere models (2), the bolometric luminosity (3), the position (optical position if counterpart is identified, otherwise X-ray position which has a typical error of $\pm 20''$) (4), the type of system (SSS = Cal 83 like supersoft source; PN = planetary nebula; Sy = symbiotic system; N = Nova) (5), the binary period (6), the mass of WD (7), the mass of the companion (8) and References (9).

| Name | Countrate[1] (cts/s) | T[2] (eV) | $L_{bol}$ (erg/s) | Type | Period |
|---|---|---|---|---|---|
| **Large Magellanic Cloud** | | | | | |
| RX J0439.8−6809 | 1.35 | 20–25 (wd) | $10-14\times 10^{37}$ | SSS | 3.37 h |
| RX J0513.9−6951 | <0.06–2.0 | 30–40 (bb) | $0.1-2\times 10^{38}$ | SSS | 18.24 h |
| RX J0527.8−6954 | 0.004–0.25 | 18–45 (bb) | $1-10\times 10^{37}$ | SSS? | |
| RX J0537.7−7034 | 0.02 | 18–30 (bb) | $0.6-2\times 10^{37}$ | | |
| CAL 83 | 0.98 | 20–50 (bb) | $1-10\times 10^{38}$ | SSS | 1.04 d |
| CAL 87 | 0.09 | 65–75 (wd) | $6-20\times 10^{37}$ | SSS | 10.6 h |
| RX J0550.0−7151 | <0.02–0.9 | 25–40 (bb) | | | |
| **Small Magellanic Cloud** | | | | | |
| 1E 0035.4−7230 | 0.33 | 40–50 (wd) | $0.8-2\times 10^{37}$ | SSS | 4.1 h |
| RX J0048.4−7332 | 0.19 | 25–45 (wd) | $1-8\times 10^{38}$ | Sy | |
| RX J0058.6−7146 | <0.001–0.7 | 15–70 (bb) | $2\times 10^{36}$ | | |
| 1E 0056.8−7154 | 0.29 | 30–40 (wd) | $2\times 10^{37}$ | PN | |
| **Andromeda Galaxy (M31)** | | | | | |
| RX J0037.4+4015 | $0.3\times 10^{-3}$ | 43 | | | |
| RX J0038.5+4014 | $0.8\times 10^{-3}$ | 45 | | | |
| RX J0038.6+4020 | $1.7\times 10^{-3}$ | 43 | | | |
| RX J0039.6+4054 | $0.4\times 10^{-3}$ | 45 | | | |
| RX J0040.4+4009 | $0.8\times 10^{-3}$ | 42 | | | |
| RX J0040.7+4015 | $1.3\times 10^{-3}$ | 42 | | | |
| RX J0041.5+4040 | $0.3\times 10^{-3}$ | 40 | | | |
| RX J0041.8+4059 | $0.5\times 10^{-3}$ | 43 | | | |
| RX J0042.4+4044 | $1.7\times 10^{-3}$ | 43 | | | |
| RX J0043.5+4207 | $2.2\times 10^{-3}$ | 45 | | | |
| RX J0044.0+4118 | $2.5\times 10^{-3}$ | 42 | | | |
| RX J0045.4+4154 | $<10^{-5}$–0.03 | 70–90 (wd) | $5-10\times 10^{37}$ | | |
| RX J0045.5+4206 | $3.1\times 10^{-3}$ | 20-48 (bb) | $7\times 10^{37}$ | | |
| RX J0046.2+4144 | $2.1\times 10^{-3}$ | 38 | | | |
| RX J0046.2+4138 | $1.1\times 10^{-3}$ | 40 | | | |
| RX J0047.6+4205 | $1.0\times 10^{-3}$ | 39 | | | |

| Name | Countrate[1] (cts/s) | T[2] (eV) | $L_{bol}$ (erg/s) | Type | Period |
|---|---|---|---|---|---|
| **Galactic Sources** | | | | | |
| RX J0019.8+2156 | 2.0 | 25–37 (wd) | $3-9 \times 10^{36}$ | SSS | 15.85 h |
| RX J0925.7−4758 | 1.0 | 70–75 (wd) | $3-7 \times 10^{35}$ [3] | SSS | 3.5-4 d |
| GQ Mus | 0.1 | 25–35 (bb) | $1-2 \times 10^{38}$ | N | 1.41 h |
| 1E 1339.8+2837 | 0.01–1.1 | 20–45 (bb) | $0.12-10 \times 10^{35}$ | | |
| AG Dra | 1.0 | 10–15 (bb) | $1.4 \times 10^{36}$ [3] | Sy | 554 d |
| RR Tel | 0.18 | 12 (wd) | $1.3 \times 10^{37}$ | Sy | 387 d |
| Nova Cyg 1992 | 0.03–76 | | $2 \times 10^{38}$ | N | 1.95 h |

[1] Countrates in the ROSAT PSPC corrected for vignetting, i.e. absorbed on-axis count rates. Count rates in the HRI have been converted to PSPC rates using a conversion factor of PSPC/HRI = 7.8 (Greiner et al. 1996a).
[2] Temperatures for the M31 sources are the maximum blackbody temperatures derived from the hardness ratios at the appropriate absorbing column (Greiner et al. 1996b).
[3] Luminosity for assumed distance of 1 kpc.

I would like to emphasise that every user of this catalog should spare no pains to consult the original papers in order to avoid propagation of my errors in the literature. I will keep this catalog updated, and would appreciate (1) being informed on any errors users might discover and (2) getting preprints on supersoft sources to be included in the next version. An electronic version of this catalog will be available on the Web shortly after this volume has appeared (http://www.rosat.mpe-garching.mpg.de/~jcg).

*Acknowledgements:* I appreciate the help of many collegues who sent preprints and reprints of their work. Special thanks to R. DiStefano for her steady encouragement to produce this catalog and for extensive discussion on its content. I apologise to anyone whose paper slipped through the literature search. JG is supported by the Deutsche Agentur für Raumfahrtangelegenheiten (DARA) GmbH under contract FKZ 50 OR 9201. The *ROSAT* project is supported by the German Bundesministerium für Bildung, Forschung, Wissenschaft und Technik (BMBW/DARA) and the Max-Planck-Society. This research has made use of the Simbad database, operated at CDS, Strasbourg, France.

## RX J0019.8+2156

**R.A.:** $00^h19^m50\overset{s}{.}0$     LII: $113°30$     Discovery: [1]     SSS
**Dec.:** $+21°56'54''$     BII: $-40°33$     Coordinates: [3]

**General Data**
   D [kpc]: 2±1 [2]     $N_H^{gal}$ [$10^{21}$ cm$^{-2}$]: 0.429
   $L_{bol}$ [erg/s]: $3-9\times 10^{36}$ [2]     Orbital Period: 15.85 h [2]
   Mass of central object [$M_\odot$]: 1.0–1.35 [4]     Mass of companion [$M_\odot$]:
        Spectral type:

**X-ray Data**
   T [eV]: 25-37 (wd)     $N_H^{fit}$ [$10^{21}$ cm$^{-2}$]: 0.4–1.1 [2]
   Orbital Modulation: quasi-sinusoidal, 10% amplitude [2]
   Variability: constant between 1990–1995 [2]

**Optical/UV Data**
   Finding Chart: [3]     $m_B$ [mag]: 12.3     $m_V$ [mag]: 12.2
   Orbital Modulation: 0.5 mag quasi-sinusoidal [2],[7]
   Opt. Spectrum: [2]     UV Spectrum: [2],[8]
   Opt. Variability: 3 different timescales: 2 hr, weeks-months, 40 yrs [2], [3]
   UV Variability: 20% irregular [8]
   Nebula [erg/s]:     Wind mass loss: 1000 km/s [2]

## 1E 0035.4−7230     RX J0037.3−7214

R.A.: $00^h37^m19\overset{s}{.}8$    LII: $304\overset{\circ}{.}45$    Discovery: [3]    SSS
Dec.: $-72°14'13''$    BII: $-44\overset{\circ}{.}85$    Coordinates: [5], [9]

### General Data
D [kpc]: SMC                                  $N_H^{gal}$ [$10^{21}$ cm$^{-2}$]: 0.649
$L_{bol}$ [erg/s]: $0.8-2\times10^{37}$ [3], [4], [10]     Orbital Period: 4.1 h [6], [9]
Mass of central object [$M_\odot$]:            Mass of companion [$M_\odot$]:
                                              Spectral type:

### X-ray Data
T [eV]: 40-50 (wd)                            $N_H^{fit}$ [$10^{21}$ cm$^{-2}$]: 0.35–0.8 [4]
Orbital Modulation: strong, nearly sinusoidal [11], [9]
Variability: amplitude of factor 3, orbital variation not excluded [4]

### Optical/UV Data
Finding Chart: [9]          $m_B$ [mag]:20.0          $m_V$ [mag]:20.2
Orbital Modulation: sinusoidal, 0.4 mag amplitude [9]
Opt. Spectrum: [9],[12]                       UV Spectrum:
Opt. Variability:
UV Variability:
Nebula [erg/s]: $<10^{34.6}$ (OIII) [8]       Wind mass loss:

### References
[1] Seward F.D., Mitchell M., 1981, ApJ 243, 736: X-ray discovery in *Einstein* data
[2] Jones L.R., Pye J.P., McHardy I.M., Fairall A.P., 1985, Space Sci. Rev. 40, 693: suggested optical counterpart (no position or finding chart)
[3] Wang Q., Wu X., 1992, ApJS 78, 391: *Einstein* SMC survey, discovery of soft and luminous X-ray spectrum
[4] Kahabka P., Pietsch W., Hasinger G., 1994, A&A 288, 538: ROSAT X-ray data on several SMC/LMC sources
[5] Orio M., Della Valle M., Massone G., Ögelman H., 1994, A&A 289, L11: X-ray position (*ROSAT* PSPC), optical ID, finding chart, optical spectrum
[6] Schmidtke P.C., Cowley A.P., McGrath T.K., Hutchings J.B., Crampton D., 1994, IAUC 6107: orbital period, colours
[7] Greiner J., 1995, (unpublished): X-ray position (*ROSAT* HRI)
[8] Remillard R.A., Rappaport S., Macri L.M., 1995, ApJ 439, 646: ionisation nebula limit
[9] Schmidtke P.C., Cowley A.P., McGrath T.K., Hutchings J.B., Crampton D., 1996, AJ 111, 788: optical data, finding chart
[10] Teeseling A. van, Heise J., Kahabka P., 1996, in Compact stars in binaries, IAU Symp. 165, Eds J. van Paradijs et al., p. 445: WD atmosphere modelling
[11] Kahabka P., 1996, A&A (in press): orbital modulation of X-rays
[12] van Teeseling A., Reinsch K., Beuermann K., Thomas H.-C., Pakull M.W., 1996, this volume p. **??**: phase-resolved spectroscopy, radial velocity

## RX J0037.4+4015

**R.A.:** $00^h37^m25{.}^s3$     LII: $120°04$      Discovery: [1]
**Dec.:** $+40°15'16''$          BII: $-22°54$      Coordinates: [1]      unidentified

**General Data**
  D [kpc]: M 31                                $N_H^{gal}$ [$10^{21}$ cm$^{-2}$]: 0.62
  $L_{bol}$ [erg/s]:                           Orbital Period:
  Mass of central object [$M_\odot$]:          Mass of companion [$M_\odot$]:
                                               Spectral type:

**X-ray Data**
  T [eV]: <43 (bb) [2]                         $N_H^{fit}$ [$10^{21}$ cm$^{-2}$]:
  Orbital Modulation:
  Variability: no [2]

**Optical/UV Data**
  Finding Chart:            $m_B$ [mag]:              $m_V$ [mag]:
  Orbital Modulation:
  Opt. Spectrum:                                UV Spectrum:
  Opt. Variability:
  UV Variability:
  Nebula [erg/s]:                               Wind mass loss:

**References**
[1] Supper R., Hasinger G., Pietsch W., Trümper J., Jain A., Magnier E.A., Lewin W.H.G., van Paradijs J., 1996, A&A (in press): ROSAT data of M31, position
[2] Greiner J., Supper R., Magnier E.A., 1996, this volume p. **??**: X-ray spectral data, position

## RX J0038.5+4014

**R.A.:** $00^h 38^m 32^s\!.1$     LII: $120°\!.27$     Discovery: [1]
**Dec.:** $+40°\, 14'\, 39''$     BII: $-22°\!.56$     Coordinates: [1]    unidentified

### General Data
D [kpc]: M 31                                  $N_H^{gal}\ [10^{21}\ cm^{-2}]$: 0.62
$L_{bol}$ [erg/s]:                              Orbital Period:
Mass of central object $[M_\odot]$:             Mass of companion $[M_\odot]$:
                                                Spectral type:

### X-ray Data
T [eV]: <45 (bb) [2]                            $N_H^{fit}\ [10^{21}\ cm^{-2}]$:
Orbital Modulation:
Variability: no [2]

### Optical/UV Data
Finding Chart:              $m_B$ [mag]:                 $m_V$ [mag]:
Orbital Modulation:
Opt. Spectrum:                                  UV Spectrum:
Opt. Variability:
UV Variability:
Nebula [erg/s]:                                 Wind mass loss:

### References
[1] Supper R., Hasinger G., Pietsch W., Trümper J., Jain A., Magnier E.A., Lewin W.H.G., van Paradijs J., 1996, A&A (in press): ROSAT data of M31, position
[2] Greiner J., Supper R., Magnier E.A., 1996, this volume p. **??**: X-ray spectral data, position

## RX J0038.6+4020

**R.A.:** $00^h38^m40\overset{s}{.}9$     **LII:** $120°30$     Discovery: [1]
**Dec.:** $+40°20'00''$     **BII:** $-22°47$     Coordinates: [1]     unidentified

**General Data**
  D [kpc]: M 31                              $N_H^{gal}$ [$10^{21}$ cm$^{-2}$]: 0.62
  $L_{bol}$ [erg/s]:                         Orbital Period:
  Mass of central object [$M_\odot$]:        Mass of companion [$M_\odot$]:
                                             Spectral type:

**X-ray Data**
  T [eV]: <43 (bb) [2]                       $N_H^{fit}$ [$10^{21}$ cm$^{-2}$]:
  Orbital Modulation:
  Variability: no [2]

**Optical/UV Data**
  Finding Chart: [2]         $m_B$ [mag]:         $m_V$ [mag]:
  Orbital Modulation:
  Opt. Spectrum:                              UV Spectrum:
  Opt. Variability:
  UV Variability:
  Nebula [erg/s]:                             Wind mass loss:

**References**
[1] Supper R., Hasinger G., Pietsch W., Trümper J., Jain A., Magnier E.A., Lewin W.H.G., van Paradijs J., 1996, A&A (in press): ROSAT data of M31, position
[2] Greiner J., Supper R., Magnier E.A., 1996, this volume p. **??**: X-ray spectral data, position, finding chart

## RX J0039.6+4054

**R.A.:** $00^h39^m38\overset{s}{.}5$    LII: $120\overset{\circ}{.}53$    Discovery: [1]
**Dec.:** $+40°54'09''$    BII: $-21\overset{\circ}{.}91$    Coordinates: [1]    unidentified

### General Data
D [kpc]: M 31    $N_H^{gal}$ [$10^{21}$ cm$^{-2}$]: 0.62
$L_{bol}$ [erg/s]:    Orbital Period:
Mass of central object [$M_\odot$]:    Mass of companion [$M_\odot$]:
   Spectral type:

### X-ray Data
T [eV]: <45 (bb) [2]    $N_H^{fit}$ [$10^{21}$ cm$^{-2}$]:
Orbital Modulation:
Variability: no [2]

### Optical/UV Data
Finding Chart: [2]    $m_B$ [mag]:    $m_V$ [mag]:
Orbital Modulation:
Opt. Spectrum:    UV Spectrum:
Opt. Variability:
UV Variability:
Nebula [erg/s]:    Wind mass loss:

### References
[1] Supper R., Hasinger G., Pietsch W., Trümper J., Jain A., Magnier E.A., Lewin W.H.G., van Paradijs J., 1996, A&A (in press): ROSAT data of M31, position
[2] Greiner J., Supper R., Magnier E.A., 1996, this volume p. **??**: X-ray spectral data, position, finding chart

## RX J0040.4+4009

**R.A.:** $00^h40^m26\overset{s}{.}3$     **LII:** $120°65$     Discovery: [1]
**Dec.:** $+40°09'01''$     **BII:** $-22°67$     Coordinates: [1]     unidentified

**General Data**
  D [kpc]: M 31                                $N_H^{gal}$ [$10^{21}$ cm$^{-2}$]: 0.62
  $L_{bol}$ [erg/s]:                           Orbital Period:
  Mass of central object [$M_\odot$]:          Mass of companion [$M_\odot$]:
                                               Spectral type:

**X-ray Data**
  T [eV]: <42 (bb) [2]                         $N_H^{fit}$ [$10^{21}$ cm$^{-2}$]:
  Orbital Modulation:
  Variability: no [2]

**Optical/UV Data**
  Finding Chart:           $m_B$ [mag]:                $m_V$ [mag]:
  Orbital Modulation:
  Opt. Spectrum:                                UV Spectrum:
  Opt. Variability:
  UV Variability:
  Nebula [erg/s]:                               Wind mass loss:

**References**
[1] Supper R., Hasinger G., Pietsch W., Trümper J., Jain A., Magnier E.A., Lewin W.H.G., van Paradijs J., 1996, A&A (in press): ROSAT data of M31, position
[2] Greiner J., Supper R., Magnier E.A., 1996, this volume p. **??**: X-ray spectral data, position

## RX J0040.7+4015

**R.A.:** $00^h 40^m 43^s.2$         **LII:** $120°.72$         Discovery: [1]
**Dec.:** $+40° 15' 18''$         **BII:** $-22°.57$         Coordinates: [1]    unidentified

### General Data
D [kpc]: M 31                                     $N_H^{gal}$ [$10^{21}$ cm$^{-2}$]: 0.62
$L_{bol}$ [erg/s]:                                Orbital Period:
Mass of central object [$M_\odot$]:               Mass of companion [$M_\odot$]:
                                                  Spectral type:

### X-ray Data
T [eV]: <42 (bb) [2]                              $N_H^{fit}$ [$10^{21}$ cm$^{-2}$]:
Orbital Modulation:
Variability: no [2]

### Optical/UV Data
Finding Chart: [2]         $m_B$ [mag]:           $m_V$ [mag]:
Orbital Modulation:
Opt. Spectrum:                                    UV Spectrum:
Opt. Variability:
UV Variability:
Nebula [erg/s]:                                   Wind mass loss:

### References
[1] Supper R., Hasinger G., Pietsch W., Trümper J., Jain A., Magnier E.A., Lewin W.H.G., van Paradijs J., 1996, A&A (in press): ROSAT data of M31, position
[2] Greiner J., Supper R., Magnier E.A., 1996, this volume p. **??**: X-ray spectral data, position, finding chart

## RX J0041.5+4040

**R.A.:** $00^h41^m30\overset{s}{.}2$  **LII:** $120\overset{\circ}{.}90$  Discovery: [1]
**Dec.:** $+40°40'04''$  **BII:** $-22\overset{\circ}{.}16$  Coordinates: [1]   unidentified

**General Data**
  D [kpc]: M 31                               $N_H^{gal}$ [$10^{21}$ cm$^{-2}$]: 0.62
  $L_{bol}$ [erg/s]:                          Orbital Period:
  Mass of central object [$M_\odot$]:         Mass of companion [$M_\odot$]:
                                              Spectral type:

**X-ray Data**
  T [eV]: <40 (bb) [2]                        $N_H^{fit}$ [$10^{21}$ cm$^{-2}$]:
  Orbital Modulation:
  Variability: no [2]

**Optical/UV Data**
  Finding Chart: [2]        $m_B$ [mag]:          $m_V$ [mag]:
  Orbital Modulation:
  Opt. Spectrum:                                  UV Spectrum:
  Opt. Variability:
  UV Variability:
  Nebula [erg/s]:                                 Wind mass loss:

**References**
[1] Supper R., Hasinger G., Pietsch W., Trümper J., Jain A., Magnier E.A., Lewin W.H.G., van Paradijs J., 1996, A&A (in press): ROSAT data of M31, position
[2] Greiner J., Supper R., Magnier E.A., 1996, this volume p. **??**: X-ray spectral data, position, finding chart

## RX J0041.8+4059

**R.A.:** $00^h 41^m 49\overset{s}{.}9$   LII: $120\overset{\circ}{.}98$   Discovery: [1]
**Dec.:** $+40°59'21''$   BII: $-21\overset{\circ}{.}85$   Coordinates: [1]   unidentified

**General Data**
  D [kpc]: M 31   $N_H^{gal}$ [$10^{21}$ cm$^{-2}$]: 0.62
  $L_{bol}$ [erg/s]:   Orbital Period:
  Mass of central object [$M_\odot$]:   Mass of companion [$M_\odot$]:
    Spectral type:

**X-ray Data**
  T [eV]: <43 (bb) [2]   $N_H^{fit}$ [$10^{21}$ cm$^{-2}$]:
  Orbital Modulation:
  Variability: no [2]

**Optical/UV Data**
  Finding Chart: [2]   $m_B$ [mag]:   $m_V$ [mag]:
  Orbital Modulation:
  Opt. Spectrum:   UV Spectrum:
  Opt. Variability:
  UV Variability:
  Nebula [erg/s]:   Wind mass loss:

**References**
[1] Supper R., Hasinger G., Pietsch W., Trümper J., Jain A., Magnier E.A., Lewin W.H.G., van Paradijs J., 1996, A&A (in press): ROSAT data of M31, position
[2] Greiner J., Supper R., Magnier E.A., 1996, this volume p. **??**: X-ray spectral data, position, finding chart

## RX J0042.4+4044

**R.A.:** $00^{\rm h}42^{\rm m}27\overset{\rm s}{.}6$     LII: $121\overset{\circ}{.}10$     Discovery: [1]
**Dec.:** $+40°44'32''$     BII: $-22\overset{\circ}{.}10$     Coordinates: [1]    unidentified

**General Data**
   D [kpc]: M 31            $N_{\rm H}^{\rm gal}$ [$10^{21}$ cm$^{-2}$]: 0.62
   $L_{\rm bol}$ [erg/s]:            Orbital Period:
   Mass of central object [$M_\odot$]:            Mass of companion [$M_\odot$]:
                                       Spectral type:

**X-ray Data**
   T [eV]: <43 (bb) [2]            $N_{\rm H}^{\rm fit}$ [$10^{21}$ cm$^{-2}$]:
   Orbital Modulation:
   Variability: no [2]

**Optical/UV Data**
   Finding Chart: [2]            $m_B$ [mag]:            $m_V$ [mag]:
   Orbital Modulation:
   Opt. Spectrum:            UV Spectrum:
   Opt. Variability:
   UV Variability:
   Nebula [erg/s]:            Wind mass loss:

**References**
[1] Supper R., Hasinger G., Pietsch W., Trümper J., Jain A., Magnier E.A., Lewin W.H.G., van Paradijs J., 1996, A&A (in press): ROSAT data of M31, position
[2] Greiner J., Supper R., Magnier E.A., 1996, this volume p. **??**: X-ray spectral data, position, finding chart

## RX J0043.5+4207

**R.A.:** $00^h43^m35\overset{s}{.}9$      **LII:** $121°31$      Discovery: [1]

**Dec.:** $+42°07'30''$      **BII:** $-22°72$      Coordinates: [1]   unidentified

### General Data
D [kpc]: M 31                                $N_H^{gal}$ [$10^{21}$ cm$^{-2}$]: 0.62
$L_{bol}$ [erg/s]:                           Orbital Period:
Mass of central object [$M_\odot$]:          Mass of companion [$M_\odot$]:
                                             Spectral type:

### X-ray Data
T [eV]: <45 (bb) [2]                         $N_H^{fit}$ [$10^{21}$ cm$^{-2}$]:
Orbital Modulation:
Variability: no [2]

### Optical/UV Data
Finding Chart:           $m_B$ [mag]:                $m_V$ [mag]:
Orbital Modulation:
Opt. Spectrum:                               UV Spectrum:
Opt. Variability:
UV Variability:
Nebula [erg/s]:                              Wind mass loss:

### References
[1] Supper R., Hasinger G., Pietsch W., Trümper J., Jain A., Magnier E.A., Lewin W.H.G., van Paradijs J., 1996, A&A (in press): ROSAT data of M31, position
[2] Greiner J., Supper R., Magnier E.A., 1996, this volume p. **??**: X-ray spectral data, position

## RX J0044.0+4118

**R.A.:** $00^{\rm h}44^{\rm m}04\overset{\rm s}{.}8$    LII: $121\overset{\circ}{.}45$    Discovery: [1]
**Dec.:** $+41°18'20''$    BII: $-21\overset{\circ}{.}54$    Coordinates: [1]    unidentified

**General Data**
  D [kpc]: M 31    $N_H^{\rm gal}$ [$10^{21}$ cm$^{-2}$]: 0.62
  $L_{\rm bol}$ [erg/s]:    Orbital Period:
  Mass of central object [$M_\odot$]:    Mass of companion [$M_\odot$]:
      Spectral type:

**X-ray Data**
  T [eV]: <42 (bb) [2]    $N_H^{\rm fit}$ [$10^{21}$ cm$^{-2}$]:
  Orbital Modulation:
  Variability: no [2]

**Optical/UV Data**
  Finding Chart: [2]    $m_B$ [mag]:    $m_V$ [mag]:
  Orbital Modulation:
  Opt. Spectrum:    UV Spectrum:
  Opt. Variability:
  UV Variability:
  Nebula [erg/s]:    Wind mass loss:

**References**
[1] Supper R., Hasinger G., Pietsch W., Trümper J., Jain A., Magnier E.A., Lewin W.H.G., van Paradijs J., 1996, A&A (in press): ROSAT data of M31, position
[2] Greiner J., Supper R., Magnier E.A., 1996, this volume p. **??**: X-ray spectral data, position, finding chart

## RX J0045.4+4154

**R.A.:** $00^h 45^m 29\overset{s}{.}0$     **LII:** $121\overset{\circ}{.}75$     Discovery: [2]
**Dec.:** $+41°54'08''$     **BII:** $-20\overset{\circ}{.}96$     Coordinates: [2]     unidentified

**General Data**
  D [kpc]: M 31                                   $N_H^{gal}$ [$10^{21}$ cm$^{-2}$]: 0.84
  $L_{bol}$ [erg/s]: 5–10$\times 10^{37}$ [2]     Orbital Period:
  Mass of central object [$M_\odot$]:             Mass of companion [$M_\odot$]:
                                                  Spectral type:

**X-ray Data**
  T [eV]: <154 (wd) [2],[3]     $N_H^{fit}$ [$10^{21}$ cm$^{-2}$]:
  Orbital Modulation:
  Variability: transient [2]

**Optical/UV Data**
  Finding Chart: [4]           $m_B$ [mag]:         $m_V$ [mag]:
  Orbital Modulation:
  Opt. Spectrum:               UV Spectrum:
  Opt. Variability:
  UV Variability:
  Nebula [erg/s]:              Wind mass loss:

**References**
[1] White N.E., Giommi P., Angelini L., Fantasia S., 1994, IAU Circ. 6064: X-ray discovery, position
[2] White N.E., Giommi P., Heise J., Angelini L., Fantasia S., 1995, ApJ 445, L125: X-ray discovery, position, atmosphere modelling of X-ray data
[3] Supper R., Hasinger G., Pietsch W., Trümper J., Jain A., Magnier E.A., Lewin W.H.G., van Paradijs J., 1996, A&A (in press): ROSAT data of M31, position
[4] Greiner J., Supper R., Magnier E.A., 1996, this volume p. **??**: X-ray spectral data, position, finding chart

## RX J0045.5+4206

**R.A.:** $00^h 45^m 32\overset{s}{.}3$  **LII:** $121\overset{\circ}{.}76$  Discovery: [1]
**Dec.:** $+42°06'59''$  **BII:** $-20\overset{\circ}{.}74$  Coordinates: [1]   unidentified

### General Data
D [kpc]: M 31  $N_H^{gal}$ [$10^{21}$ cm$^{-2}$]: 0.84
$L_{bol}$ [erg/s]: $7 \times 10^{37}$ [2]  Orbital Period:
Mass of central object [$M_\odot$]:  Mass of companion [$M_\odot$]:
  Spectral type:

### X-ray Data
T [eV]: 20–48 (bb) [2]  $N_H^{fit}$ [$10^{21}$ cm$^{-2}$]:
Orbital Modulation:
Variability: no [2]

### Optical/UV Data
Finding Chart: [2]  $m_B$ [mag]:  $m_V$ [mag]:
Orbital Modulation:
Opt. Spectrum:  UV Spectrum:
Opt. Variability:
UV Variability:
Nebula [erg/s]:  Wind mass loss:

### References
[1] Supper R., Hasinger G., Pietsch W., Trümper J., Jain A., Magnier E.A., Lewin W.H.G., van Paradijs J., 1996, A&A (in press): ROSAT data of M31, position
[2] Greiner J., Supper R., Magnier E.A., 1996, this volume p. **??**: X-ray spectral data, position, finding chart

## RX J0046.2+4144

**R.A.:** $00^h46^m15\overset{s}{.}6$     LII: $121\overset{\circ}{.}90$     Discovery: [1]
**Dec.:** $+41°44'36''$     BII: $-21\overset{\circ}{.}12$     Coordinates: [1]     unidentified

**General Data**
  D [kpc]: M 31                              $N_H^{gal}$ [$10^{21}$ cm$^{-2}$]: 0.84
  $L_{bol}$ [erg/s]:                          Orbital Period:
  Mass of central object [$M_\odot$]:         Mass of companion [$M_\odot$]:
                                              Spectral type:

**X-ray Data**
  T [eV]: <38 (bb) [2]                        $N_H^{fit}$ [$10^{21}$ cm$^{-2}$]:
  Orbital Modulation:
  Variability: no [2]

**Optical/UV Data**
  Finding Chart: [2]          $m_B$ [mag]:                  $m_V$ [mag]:
  Orbital Modulation:
  Opt. Spectrum:                              UV Spectrum:
  Opt. Variability:
  UV Variability:
  Nebula [erg/s]:                             Wind mass loss:

**References**
[1] Supper R., Hasinger G., Pietsch W., Trümper J., Jain A., Magnier E.A., Lewin W.H.G., van Paradijs J., 1996, A&A (in press): ROSAT data of M31, position
[2] Greiner J., Supper R., Magnier E.A., 1996, this volume p. **??**: X-ray spectral data, position, finding chart

## RX J0046.2+4138

**R.A.:** $00^h 46^m 17\overset{s}{.}8$     **LII:** $121\overset{\circ}{.}90$     Discovery: [1]
**Dec.:** $+41°38'48''$     **BII:** $-21\overset{\circ}{.}21$     Coordinates: [1]    unidentified

**General Data**
 D [kpc]: M 31     $N_H^{gal}$ [$10^{21}$ cm$^{-2}$]: 0.84
 $L_{bol}$ [erg/s]:     Orbital Period:
 Mass of central object [$M_\odot$]:     Mass of companion [$M_\odot$]:
     Spectral type:

**X-ray Data**
 T [eV]: <40 (bb) [2]     $N_H^{fit}$ [$10^{21}$ cm$^{-2}$]:
 Orbital Modulation:
 Variability: no [2]

**Optical/UV Data**
 Finding Chart:     $m_B$ [mag]:     $m_V$ [mag]:
 Orbital Modulation:
 Opt. Spectrum:     UV Spectrum:
 Opt. Variability:
 UV Variability:
 Nebula [erg/s]:     Wind mass loss:

**References**
[1] Supper R., Hasinger G., Pietsch W., Trümper J., Jain A., Magnier E.A., Lewin W.H.G., van Paradijs J., 1996, A&A (in press): ROSAT data of M31, position
[2] Greiner J., Supper R., Magnier E.A., 1996, this volume p. **??**: X-ray spectral data, position

## RX J0047.6+4205

**R.A.:** $00^h 47^m 38\overset{s}{.}5$    **LII:** $122\overset{\circ}{.}18$    Discovery: [1]
**Dec.:** $+42°05'07''$    **BII:** $-20\overset{\circ}{.}78$    Coordinates: [1]    unidentified

### General Data
D [kpc]: M 31     $N_H^{gal}$ [$10^{21}$ cm$^{-2}$]: 0.84
$L_{bol}$ [erg/s]:     Orbital Period:
Mass of central object [$M_\odot$]:     Mass of companion [$M_\odot$]:
    Spectral type:

### X-ray Data
T [eV]: <39 (bb) [2]     $N_H^{fit}$ [$10^{21}$ cm$^{-2}$]:
Orbital Modulation:
Variability: no [2]

### Optical/UV Data
Finding Chart:     $m_B$ [mag]:     $m_V$ [mag]:
Orbital Modulation:
Opt. Spectrum:     UV Spectrum:
Opt. Variability:
UV Variability:
Nebula [erg/s]:     Wind mass loss:

### References
[1] Supper R., Hasinger G., Pietsch W., Trümper J., Jain A., Magnier E.A., Lewin W.H.G., van Paradijs J., 1996, A&A (in press): ROSAT data of M31, position
[2] Greiner J., Supper R., Magnier E.A., 1996, this volume p. **??**: X-ray spectral data, position

## RX J0048.4−7332                                                        SMC 3

**R.A.:** $00^h48^m20^s.8$     LII: $303°23$     Discovery: [2]     **Sy**
**Dec.:** $-73°31'53''$     BII: $-43°59$     Coordinates: [1]

**General Data**
  D [kpc]: SMC     $N_H^{gal}$ [$10^{21}$ cm$^{-2}$]: 0.516
  $L_{bol}$ [erg/s]: $1-8 \times 10^{38}$ [5], [6]     Orbital Period:
  Mass of central object [$M_\odot$]:     Mass of companion [$M_\odot$]:
      Spectral type:

**X-ray Data**
  T [eV]: 25–45 (wd) [2], [6], [5]     $N_H^{fit}$ [$10^{21}$ cm$^{-2}$]: 1.2–7.0 [2]
  Orbital Modulation:
  Variability:

**Optical/UV Data**
  Finding Chart: [1]     $m_B$ [mag]:     $m_V$ [mag]: 15.5
  Orbital Modulation:
  Opt. Spectrum: [1],[7]     UV Spectrum:
  Opt. Variability: 1.5 mag outburst in 1981 [1]
  UV Variability:
  Nebula [erg/s]: $<10^{34.6}$ (OIII) [4]     Wind mass loss:

**References**
[1] Morgan D.H., 1992, MNRAS 258, 639: symbiotic nature and outburst, finding chart
[2] Kahabka P., Pietsch W., Hasinger G., 1994, A&A 288, 538: ROSAT X-ray data on several SMC/LMC sources
[3] Vogel M., Morgan D.H., 1994, A&A 288, 842: IUE spectrum
[4] Remillard R.A., Rappaport S., Macri L.M., 1995, ApJ 439, 646: ionisation nebula limit
[5] Teeseling A. van, Heise J., Kahabka P., 1996, in Compact stars in binaries, IAU Symp. 165, Eds J. van Paradijs et al., p. 445: WD atmosphere modelling
[6] Jordan S., Schmutz W., Wolff B., Werner K., Mürset U., 1996, A&A (in press): model atmosphere for optical, UV and X-ray emission including luminosity, mass loss and abundances
[7] Mürset U., Schild H., Vogel M., 1996, A&A (in press): optical spectrum

## RX J0058.6−7146

**R.A.**: $00^h58^m35\overset{s}{.}8$     LII: $302\overset{\circ}{.}14$     Discovery: [1]
**Dec.**: $-71°46'02''$     BII: $-45\overset{\circ}{.}35$     Coordinates: [1]     unidentified

**General Data**
  D [kpc]: SMC     $N_H^{gal}$ [$10^{21}$ cm$^{-2}$]: 0.748
  $L_{bol}$ [erg/s]: $2\times10^{36}$     Orbital Period:
  Mass of central object [$M_\odot$]:     Mass of companion [$M_\odot$]:
       Spectral type:

**X-ray Data**
  T [eV]: 15-70 (bb)[1]     $N_H^{fit}$ [$10^{21}$ cm$^{-2}$]: 0.3–1.5 [1]
  Orbital Modulation:
  Variability: turn on within 2 days [1]

**Optical/UV Data**
  Finding Chart:     $m_B$ [mag]:     $m_V$ [mag]:
  Orbital Modulation:
  Opt. Spectrum:     UV Spectrum:
  Opt. Variability:
  UV Variability:
  Nebula [erg/s]: $<10^{34.6}$ (OIII) [2]     Wind mass loss:

**References**
[1] Kahabka P., Pietsch W., Hasinger G., 1994, A&A 288, 538: ROSAT X-ray data on several SMC/LMC sources, X-ray turn on
[2] Remillard R.A., Rappaport S., Macri L.M., 1995, ApJ 439, 646: ionisation nebula limit

## 1E 0056.8–7154                                                                    SMC N67

R.A.: $00^h58^m37\overset{s}{.}0$     LII: $302\overset{\circ}{.}12$     Discovery: [7]     PN
Dec.: $-71°35'48''$     BII: $-45\overset{\circ}{.}52$     Coordinates: [2]

**General Data**
  D [kpc]: SMC                                             $N_H^{gal}$ [$10^{21}$ cm$^{-2}$]: 0.507
  $L_{bol}$ [erg/s]: $2\times10^{37}$ [11]                 Orbital Period:
  Mass of central object [$M_\odot$]: 0.9                  Mass of companion [$M_\odot$]:
                                                           Spectral type:

**X-ray Data**
  T [eV]: 30-40 (wd) [11]                                  $N_H^{fit}$ [$10^{21}$ cm$^{-2}$]: 0.5 [11]
  Orbital Modulation:
  Variability: <20–40% [10], [9]

**Optical/UV Data**
  Finding Chart:          $m_B$ [mag]:           $m_V$ [mag]: 16.7
  Orbital Modulation:
  Opt. Spectrum:                                 UV Spectrum:
  Opt. Variability:
  UV Variability:
  Nebula [erg/s]: <$10^{34.6}$ (OIII) [3], [12]  Wind mass loss:

**References**
[1] Henize K.G., 1956, ApJS 2, 315: position
[2] Aller L.H., Keyes C.D., 1987, ApJ 320, 159: position, UV spectrum
[3] Wood P.R., Meatheringham S.J., Dopita M.A., Morgan D.H., 1987, ApJ 320, 178: optical diameter limit from Speckle imaging, OIII and H$\beta$ flux
[4] Dopita M.A., Meatheringham S.J., 1991, ApJ 367, 115: photoionisation modelling of optical spectrum, nebular parameters (L, $T_{eff}$, $R_{in}$, $R_{out}$, $M_{neb}$, density)
[5] Meatheringham S.J., Dopita M.A., 1991, ApJ Suppl. 75, 407: optical line intensities, abundances, comparison with other PN
[6] Wang Q., 1991, MNRAS 252, 47p: Einstein X-ray data, spectrum, luminosity
[7] Wang Q., Wu X., 1992, ApJS 78, 391: *Einstein* SMC survey, discovery of soft and luminous X-ray spectrum
[8] Brown T., Cordova F., Ciardullo R., Thompson R., Bond H., 1994, ApJ 422, 118: reanalysis Einstein data
[9] Kahabka P., Pietsch W., Hasinger G., 1994, A&A 288, 538: ROSAT X-ray data on several SMC/LMC sources
[10] Hughes J.P., 1994, ApJ 427, L25: variability ROSAT/Einstein
[11] Heise J., van Teeseling A., Kahabka P., 1994, A&A 288, L45
[12] Remillard R.A., Rappaport S., Macri L.M., 1995, ApJ 439, 646: ionisation nebula limit
[13] Teeseling A. van, Heise J., Kahabka P., 1996, in Compact stars in binaries, IAU Symp. 165, Eds J. van Paradijs et al., p. 445: WD atmosphere modelling

## RX J0439.8−6809

**R.A.:** $04^h39^m49^s\!.6$      **LII:** $279°\!.87$      Discovery: [1]      **SSS**
**Dec.:** $-68°09'02''$      **BII:** $-37°\!.10$      Coordinates: [3],[5],[7]

**General Data**
  D [kpc]: LMC        $N_H^{gal}$ [$10^{21}$ cm$^{-2}$]: 0.447
  $L_{bol}$ [erg/s]: $10-14 \times 10^{37}$        Orbital Period: 3.37 h [6]
  Mass of central object [$M_\odot$]:        Mass of companion [$M_\odot$]:
                                             Spectral type:

**X-ray Data**
  T [eV]: 20-25 (wd) [7]        $N_H^{fit}$ [$10^{21}$ cm$^{-2}$]: 0.25−0.4 [7]
  Orbital Modulation:
  Variability: constant 1990−1994 [1]

**Optical/UV Data**
  Finding Chart: [1],[5],[6],[7]      $m_B$ [mag]: 21.5      $m_V$ [mag]: 21.7
  Orbital Modulation: 0.15 mag [6]
  Opt. Spectrum: [5],[7]        UV Spectrum:
  Opt. Variability: marginal [5],[6],[7]
  UV Variability:
  Nebula [erg/s]: $<10^{34.6}$ (OIII) [2]        Wind mass loss:

**References**
[1] Greiner J., Hasinger G., Thomas H.-C., 1994, A&A 281, L61
[2] Remillard R.A., Rappaport S., Macri L.M., 1995, ApJ 439, 646: ionisation nebula limit
[3] Schmidtke P.C., Cowley A.P., 1995, IAU Circ. 6278: optical counterpart proposed (position and colour)
[4] Reinsch K., van Teeseling A., Beuermann K., Thomas H.-C., 1996, in Röntgenstrahlung from the Universe, Eds. H.U. Zimmermann et al. , p. 183: optical counterpart proposed (colour)
[5] van Teeseling A., Reinsch K., Beuermann K., Thomas H.-C., Pakull M.W., 1996, this volume p. **??**: HRI X-ray position, optical identification, U + V finding chart, optical spectrum, short-term variability
[6] Schmidtke P.C., Cowley A.P., 1996, this volume p. **??**: U + V finding chart, orbital period
[7] van Teeseling A., Reinsch K., Beuermann K., 1996, A&A (in press): HRI X-ray position, optical identification, U + V finding chart, optical spectrum, short-term variability

## RX J0513.9−6951                                                      HV 5682

**R.A.:** $05^h13^m50\rlap{.}^s8$    LII: $280\rlap{.}°80$       Discovery: [1]          SSS
**Dec.:** $-69°51'47''$              BII: $-33\rlap{.}°69$       Coordinates: [1], [3], [7]

**General Data**
  D [kpc]: LMC                              $N_H^{gal}$ [$10^{21}$ cm$^{-2}$]: 0.838
  $L_{bol}$ [erg/s]: 0.1–6×$10^{38}$ [1]    Orbital Period: 0.76 d [7],[10],[13]
  Mass of central object [$M_\odot$]: 1.3–1.4 [6]   Mass of companion [$M_\odot$]: >0.8 [7]
                                            Spectral type:

**X-ray Data**
  T [eV]: 30-40 (bb) [1]                    $N_H^{fit}$ [$10^{21}$ cm$^{-2}$]: 0.9 [1]
  Orbital Modulation: no
  Variability: transient with on/off time scale of 4 weeks [1],[9]

**Optical/UV Data**
  Finding Chart: [2], [3]        $m_B$ [mag]: 16.6        $m_V$ [mag]: 16.7
  Orbital Modulation: 0.1 mag [7],[10],[13]
  Opt. Spectrum: [2],[3],[7],[10],[11],[12]      UV Spectrum: [2]
  Opt. Variability: semi-regular 1 mag drops of ≈30 d duration [2],[7],[10],[11],[12]
  UV Variability:
  Nebula [erg/s]: <$10^{34.6}$ (OIII) [4]        Wind mass loss: 3800 km/s [2], [7]

**References**
[1] Schaeidt S., Hasinger G., Trümper J., 1993, A&A 270, L9: *ROSAT* X-ray data
[2] Pakull M.W., Motch C., Bianchi L., Thomas H.-C., Guibert J., Beaulieu J.-P., Grison P., Schaeidt S., 1993, A&A 278, L39: opt. ID, finding chart, opt. + UV spectrum
[3] Cowley A.P., Schmidtke P.C., Hutchings J.B., Crampton D., McGrath T.K., 1993, ApJ 418, L63: finding chart, opt. spectrum
[4] Remillard R.A., Rappaport S., Macri L.M., 1995, ApJ 439, 646: ionisation nebula limit
[5] Greiner J., 1995, Abano-Padova Conf. on Cataclysmic variables, eds. A. Bianchini, M. Della Valle, M. Orio, ASSL 205, 443: X-ray and optical variability
[6] Kahabka P., 1995, A&A 304, 227: WD mass from variability timescale
[7] Crampton D., Hutchings J.B., Cowley A.P., Schmidtke P.C., McGrath T.K., O'-Donoghue D., Harrop-Allin M.K., 1996, ApJ 456, 320: opt. photometry and spectroscopy, orbit, period, bipolar outflows
[8] Cowley A.P., Schmidtke P.C., Crampton D., Hutchings J.B., 1996, in Compact stars in binaries, IAU Symp. 165, Eds J. van Paradijs et al., p.439 : prelim. period
[9] Schaeidt S., 1996, this volume p. **??**: ROSAT HRI monitoring, several on-states
[10] Southwell K.A., Livio M., Charles P.A., Sutherland W., Alcock C., Allsman R.A., Alves D., Axelrod T.S., Bennett D.P., Cook K.H., Freeman K.C., Griest K., Guern J., Lehner M.J., Marshall S.L., Peterson B.A., Pratt M.R., Quinn P.J., Rodgers A.W., Stubbs C.W., Welch D.L., 1996, this volume p. **??**: 3 yrs optical lightcurve, semi-regular brightness drops, orbital period
[11] Reinsch K., van Teeseling A., Beuermann K., Thomas H.-C., 1996, this volume p. **??**: photometric monitoring, optical drop
[12] Reinsch K., van Teeseling A., Beuermann K., Abbott T.M.C., 1996, A&A (in press): photometric monitoring, optical drop
[13] Motch C., Pakull M.W., 1996, this volume p. **??**: orbital period

## RX J0527.8−6954

| | | |
|---|---|---|
| **R.A.:** $05^h27^m49\overset{s}{.}9$ | **LII:** $280\overset{\circ}{.}56$ | Discovery: [1],[2] |
| **Dec.:** $-69°54'09''$ | **BII:** $-32\overset{\circ}{.}50$ | Coordinates: [9]   unidentified |

**General Data**
  D [kpc]:                                    $N_H^{gal}$ [$10^{21}$ cm$^{-2}$]: 0.622
  $L_{bol}$ [erg/s]: $1-10\times10^{37}$       Orbital Period:
  Mass of central object [$M_\odot$]: 1.1–1.35 [8],[10]   Mass of companion [$M_\odot$]:
                                               Spectral type:

**X-ray Data**
  T [eV]: 18-45 (bb) [2]                       $N_H^{fit}$ [$10^{21}$ cm$^{-2}$]: 0.7–1.0 [2]
  Orbital Modulation:
  Variability: steady decrease with 5 yr timescale, possibly periodic [9],[10]

**Optical/UV Data**
  Finding Chart: [4],[10]        $m_B$ [mag]: >19         $m_V$ [mag]:
  Orbital Modulation:
  Opt. Spectrum:                                UV Spectrum:
  Opt. Variability:
  UV Variability:
  Nebula [erg/s]: $<10^{34.6}$ (OIII) [6]        Wind mass loss:

**References**
[1] Trümper J., Hasinger G., Aschenbach B., Bräuninger H., Briel U.G., Burkert W., Fink H., Pfeffermann E., Pietsch W., Predehl P., Schmitt J.H.M.M., Voges W., Zimmermann U., Beuermann K., 1991, Nat 349, 579: first report on X-ray discovery
[2] Greiner J., Hasinger G., Kahabka P. 1991, A&A 246, L17: X-ray discovery, position, temperature
[3] Orio M., Ögelman H., 1993, A&A 273, L56: X-ray fading
[4] Cowley A.P., Schmidtke P.C., Hutchings J.B., Crampton D., McGrath T.K., 1993, ApJ 418, L63: finding chart, X-ray position
[5] Hasinger G., 1994, Reviews in Modern Astronomy 7, 129: prel. X-ray lightcurve
[6] Remillard R.A., Rappaport S., Macri L.M., 1995, ApJ 439, 646: ionisation nebula limit
[7] Greiner J., 1995, in Proc. of the Abano-Terme Conference on Cataclysmic Variables, eds. A. Bianchini, M. Della Valle, M. Orio, Kluwer, ASSL 205, p. 443: prel. X-ray lightcurve
[8] Kahabka P., 1995, A&A 304, 227: WD mass from variability timescale
[9] Greiner J., Schwarz R., Hasinger G., Orio M., 1996, A&A (in press): X-ray lightcurve, improved X-ray position (HRI)
[10] Greiner J., Schwarz R., Hasinger G., Orio M., 1996, this volume p. ??: X-ray lightcurve, improved X-ray position (HRI), finding chart

## RX J0537.7−7034                                    RX J0537.6−7033

**R.A.:** $05^h 37^m 43^s\!.0$      LII: $281°\!.19$        Discovery: [1]
**Dec.:** $-70°34'15''$       BII: $-31°\!.57$        Coordinates: [1]    unidentified

**General Data**
  D [kpc]:                                          $N_H^{gal}$ [$10^{21}$ cm$^{-2}$]: 0.637
  $L_{bol}$ [erg/s]: $0.6-2 \times 10^{37}$ [1]     Orbital Period:
  Mass of central object [$M_\odot$]:               Mass of companion [$M_\odot$]:
                                                    Spectral type:

**X-ray Data**
  T [eV]: 18-30 (bb) [1]                            $N_H^{fit}$ [$10^{21}$ cm$^{-2}$]:
  Orbital Modulation:
  Variability: amplitude of factor 10 [2]

**Optical/UV Data**
  Finding Chart:                 $m_B$ [mag]:                 $m_V$ [mag]:
  Orbital Modulation:
  Opt. Spectrum:                                   UV Spectrum:
  Opt. Variability:
  UV Variability:
  Nebula [erg/s]:                                  Wind mass loss:

**References**
[1] Orio M., Ögelman H., 1993, A&A 273, L56: X-ray discovery
[2] Orio M., Della Valle M., Massone G., Ögelman H., 1996, in Proc. of Workshop on Cataclysmic Variables, Keele, June 1995 (in press): X-ray variability, optical counterpart suggested

# CAL 83                                                               LHG 83

**R.A.:** $05^h 43^m 33^s\!.5$      LII: $278°\!.56$      Discovery: [1]      SSS
**Dec.:** $-68°22'23''$      BII: $-31°\!.31$      Coordinates: [2]

**General Data**
D [kpc]: LMC                                      $N_H^{gal}$ [$10^{21}$ cm$^{-2}$]: 0.652
$L_{bol}$ [erg/s]: $1$–$10 \times 10^{38}$ [7]         Orbital Period: 1.04 d [5]
Mass of central object [$M_\odot$]:           Mass of companion [$M_\odot$]:
                                                         Spectral type:

**X-ray Data**
T [eV]: 20-50 (bb)                           $N_H^{fit}$ [$10^{21}$ cm$^{-2}$]: 0.7–0.85 [7]
Orbital Modulation: no
Variability:

**Optical/UV Data**
Finding Chart: [5]           $m_B$ [mag]: 16.2–17.3      $m_V$ [mag]: 16.2–17.3
Orbital Modulation: 0.22 mag sinusoidal [5]
Opt. Spectrum: [3]                           UV Spectrum: [3],[4]
Opt. Variability: erratic
UV Variability: 50% irregular [3],[4]
Nebula [erg/s]: OIII ($10^{35.6}$), H$\alpha$ ($10^{35.4}$) [6][9]   Wind mass loss:

**References**
[1] Long K.S., Helfand D.J., Grabelsky D.A., 1981, ApJ 248, 925: X-ray discovery (Einstein)
[2] Pakull M.W., Ilovaisky S.A., Chevalier C., 1985, Space Sci Rev 40, 229: optical identification
[3] Crampton D., Cowley A.P., Hutchings J.B., Schmidtke P.C., Thompson I.B., Liebert J., 1987, ApJ 321, 745: optical and IUE spectra, orbital period
[4] Bianchi L., Pakull M.W., 1988, in A decade of UV Astronomy with IUE, ESA SP-281, vol. 1, p. 145: UV spectrum
[5] Smale A.P., Corbet R.H., Charles P.Q., Ilovaisky S.A., Mason K.O., Motch C., Mukai K., Naylor T., Parmer A.N., van der Klis M., van Paradijs J., 1988, MNRAS 233, 51: orbital period, finding chart, optical spectra
[6] Pakull M.W., Motch C., 1989, in Extranuclear Activity in Galaxies, ed. E.J.A. Meurs, R.A.E. Fosbury, (Garching, ESO), p. 285: nebula
[7] Greiner J., Hasinger G., Kahabka P., 1991, A&A 246, L17: ROSAT X-ray spectrum and luminosity, position, temperature
[8] Brown T., Cordova F., Ciardullo R., Thompson R., Bond H., 1994, ApJ 422, 118: reanalysis Einstein data
[9] Remillard R.A., Rappaport S., Macri L.M., 1995, ApJ 439, 646: ionisation nebula

# CAL 87                                                                    LHG 87

**R.A.:** $05^h46^m52^s3$     LII: $281°75$          Discovery: [1]              SSS
**Dec.:** $-71°08'38''$       BII: $-30°76$          Coordinates:

### General Data
D [kpc]: LMC                                         $N_H^{gal}$ [$10^{21}$ cm$^{-2}$]: 0.749
$L_{bol}$ [erg/s]: $6-20 \times 10^{37}$ [12]        Orbital Period: 10.6 h [6]
Mass of central object [$M_\odot$]:                   Mass of companion [$M_\odot$]:
                                                     Spectral type:

### X-ray Data
T [eV]: 65-75 (wd) [12]                              $N_H^{fit}$ [$10^{21}$ cm$^{-2}$]:
Orbital Modulation: eclipse with amplitude of factor 3 [8]
Variability:

### Optical/UV Data
Finding Chart: [6]              $m_B$ [mag]: 19.0        $m_V$ [mag]: 18.9
Orbital Modulation: 2 mag eclipse with broad wings [5], [6]
Opt. Spectrum: [5]                                   UV Spectrum: [11]
Opt. Variability:
UV Variability:
Nebula [erg/s]: $<10^{34.6}$ (OIII) [10]             Wind mass loss:

### References
[1] Long K.S., Helfand D.J., Grabelsky D.A., 1981, ApJ 248, 925: X-ray discovery
[2] Pakull M.W., Beuermann K., Angebault L.P., Bianchi L., 1987, Ap&SS 131, 689: optical identification
[3] Pakull M.W., Beuermann K., van der Klis M., van Paradijs J., 1988, A&A 203, L27: eclipse lightcurve
[4] Callanan P.J., Machin G., Naylor T., Charles P.A., 1989, MNRAS 241, 37p: eclipse lightcurve, orbital period
[5] Cowley A.P., Schmidtke P.C., Crampton D., Hutchings J.B., 1990, ApJ 350, 288: orbital period, optical spectra
[6] Schmidtke P.C., McGrath T.K., Cowley A.P., Frattare L.M., 1993, PASP 105, 863: orbital eclipse lightcurve, finding chart, X-ray lightcurve
[7] Brown T., Cordova F., Ciardullo R., Thompson R., Bond H., 1994, ApJ 422, 118: reanalysis Einstein data
[8] Kahabka P., Pietsch W., Hasinger G., 1994, A&A 288, 538: ROSAT X-ray data on several SMC/LMC sources, X-ray eclipse in CAL 87
[9] Khruzina T.S., Cherepashchuk A.M., 1994, Astron. Zh. 71, 442: binary parameters
[10] Remillard R.A., Rappaport S., Macri L.M., 1995, ApJ 439, 646: ionisation nebula limit
[11] Hutchings J.B., Cowley A.P., Schmidtke P.C., Crampton D., 1995, AJ 110, 2394: UV spectra
[12] Teeseling A. van, Heise J., Kahabka P., 1996, in Compact stars in binaries, IAU Symp. 165, Eds J. van Paradijs et al., p. 445: WD atmosphere modelling
[13] Schandl S., Meyer-Hofmeister E., Meyer F., 1996, this volume p. **??**: modelling of orbital lightcurve, inclination
[14] Schandl S., Meyer-Hofmeister E., Meyer F., 1996, A&A (subm.): modelling of orbital lightcurve, inclination

## RX J0550.0−7151

**R.A.:** $05^h50^m00^s\!.2$         **LII:** $282°\!.52$         Discovery: [1]
**Dec.:** $-71°52'09''$         **BII:** $-30°\!.47$         Coordinates: [6]    unidentified

**General Data**
  D [kpc]:                                          $N_H^{gal}$ [$10^{21}$ cm$^{-2}$]: 0.892
  $L_{bol}$ [erg/s]:                                Orbital Period:
  Mass of central object [$M_\odot$]:               Mass of companion [$M_\odot$]:
                                                    Spectral type:

**X-ray Data**
  T [eV]: 25–40 (bb) [3],[6]                        $N_H^{fit}$ [$10^{21}$ cm$^{-2}$]: 2.0 [3]
  Orbital Modulation:
  Variability: turn-off [6]

**Optical/UV Data**
  Finding Chart: [6]         $m_B$ [mag]:           $m_V$ [mag]: >19.5
  Orbital Modulation:
  Opt. Spectrum:                                    UV Spectrum:
  Opt. Variability:
  UV Variability:
  Nebula [erg/s]: <$10^{34.6}$ (OIII) [2]           Wind mass loss:

**References**
[1] Cowley A.P., Schmidtke P.C., Hutchings J.B., Crampton D., McGrath T.K., 1993, ApJ 418, L63: X-ray discovery, position, temperature
[2] Remillard R.A., Rappaport S., Macri L.M., 1995, ApJ 439, 646: ionisation nebula limit
[3] Schmidtke P.C., Cowley A.P., 1995, IAU Circ. 6278: ROSAT HRI position in off-state, optical position and colour
[4] Charles P.A., Southwell K.A., 1996, IAU Circ. 6305: possible symbiotic ID
[5] Schmidtke P.C., Cowley A.P., 1996, this volume p. ??: difference to $2'\!.9$ nearby source RX J0549.8–7150
[6] Reinsch K., van Teeseling A., Beuermann K., Thomas H.-C., 1996, this volume p. ??: X-ray turn off, different source in $2'\!.9$ distance during off-state pointing = dMe star which is coincident with optical star from [3],[4]

## RX J0925.7−4758

R.A.: $09^h25^m46\overset{s}{.}2$     LII: $271\overset{\circ}{.}36$     Discovery: [1]     SSS
Dec.: $-47°58'17''$     BII: $1\overset{\circ}{.}88$     Coordinates: [1]

**General Data**
D [kpc]: 0.4–2 [1]     $N_H^{gal}$ [$10^{21}$ cm$^{-2}$]: 13.0
$L_{bol}$ [erg/s]: 3–7×$10^{35}$ (D/1 kpc)$^2$ [4]     Orbital Period: 3.55–4.03 d [1], [3]
Mass of central object [$M_\odot$]:     Mass of companion [$M_\odot$]:
     Spectral type:

**X-ray Data**
T [eV]: 70-75 (wd) [4]     $N_H^{fit}$ [$10^{21}$ cm$^{-2}$]: 20–24 [4]
Orbital Modulation: possible [3]
Variability: less than 50% [1]

**Optical/UV Data**
Finding Chart: [1]     $m_B$ [mag]: 19.2     $m_V$ [mag]: 17.2
Orbital Modulation: 0.3 mag amplitude
Opt. Spectrum: [1]     UV Spectrum:
Opt. Variability:
UV Variability:
Nebula [erg/s]:     Wind mass loss:

**References**
[1] Motch C., Hasinger G., Pietsch W., 1994, A&A 284, 827: X-ray discovery, optical ID, optical spectrum, prel. period, distance, finding chart
[2] Ebisawa K., Asai K., Dotani T., Mukai K., Smale A., 1996, in Röntgenstrahlung from the Universe, Eds. H.U. Zimmermann et al., MPE Report 263, p. 133: ASCA observation
[3] Motch C., 1996, this volume p. **??**: optical photometry and spectroscopy, orbital period, ROSAT HRI monitoring
[4] Hartmann H.W., Heise J., 1996, this volume p. **??**: high-gravity NLTE modelling

## GQ Mus                                              Nova Muscae 1983

**R.A.:** $11^h52^m02\overset{s}{.}5$     LII: $297°\!\!.21$      Discovery: [6],[8]            N
**Dec.:** $-67°12'24''$                   BII: $-5°\!\!.00$       Coordinates: [2]

### General Data
D [kpc]: $4.7\pm1.5$                      $N_H^{gal}$ [$10^{21}$ cm$^{-2}$]: 4.24
$L_{bol}$ [erg/s]: $1$–$2\times10^{38}$ [8]    Orbital Period: 1.41 [7]
Mass of central object [$M_\odot$]: $<1.1$–1.25 [13]    Mass of companion [$M_\odot$]: $<0.2$ [8]
                                          Spectral type:

### X-ray Data
T [eV]: 25–35 (bb) [8]                    $N_H^{fit}$ [$10^{21}$ cm$^{-2}$]: 1.0–3.4 [8]
Orbital Modulation: quasi-sinusoidal, 50% amplitude [14]
Variability: decline by factor $>30$ [12]

### Optical/UV Data
Finding Chart: [3], [4]        $m_B$ [mag]: 7–21         $m_V$ [mag]: 7–21
Orbital Modulation: 0.2 mag [7]
Opt. Spectrum: [4],[7],[9],[11]           UV Spectrum: [4]
Opt. Variability: nova outburst and decline
UV Variability:
Nebula [erg/s]:                           Wind mass loss:

### References
[1] Liller W., 1983, IAUC 3764: optical discovery, coordinate
[2] Cragg T., Nikoloff I., Johnston J., 1983, IAUC 3766: improved coordinate
[3] Bateson, Morel, 1983, Charts Southern Var. 16: finding chart
[4] Krautter J., Beuermann K., Leitherer C., Oliva E., Moorwood A.F.M., Deul E., Wargau W., Klare G., Kohoutek L., Paradijs J. van, Wolf B., 1984, A&A 137, 307: finding chart, IR, optical and UV spectra, distance, luminosity
[5] Whitelock P.A., Carter B.S., Feast M.W., Glass I.S., Laney D., Menzies J.W., Walsh J., Williams P.M., 1984, MNRAS 211, 421: optical + IR photometry/spectroscopy
[6] Ögelman H., Beuermann K., Krautter J., 1984, ApJ 287, L31: EXOSAT X-ray detection 1 yr after maximum
[7] Diaz M.P., Steiner J.E., 1989, ApJ 339, L41: orbital period, optical spectrum
[8] Ögelman H., Orio M., Krautter J., Starrfield S., 1993, Nat. 361, 331: X-ray discovery (ROSAT)
[9] Pequignot D., Petitjean P., Boisson C., Krautter J., 1993, A&A 271, 219: optical spectra from 1984–1988
[10] Diaz M.P., Steiner J.E., 1994, ApJ 425, 252: magnetic nature, photometry, spectroscopy, Doppler tomography, accretion rate, distance
[11] Diaz M.P., Williams R.E., Phillips M.M., Hamuy M., 1995, MNRAS 277, 959: optical spectra, photometry, photoionisation modelling
[12] Shanley L., Ögelman H., Gallagher J.S., Orio M., Krautter J., 1995, ApJ 438, L95: X-ray decline
[13] Kahabka P., 1995, A&A 304, 227: WD mass from variability timescale
[14] Kahabka P., 1996, A&A (in press): orbital modulation of X-rays

## 1E 1339.8+2837                                              RX J1342.1+2822

**R.A.:** $13^{\mathrm{h}}42^{\mathrm{m}}09\overset{\mathrm{s}}{.}8$     **LII:** $42°23$     Discovery: [3]
**Dec.:** $28°22'45''$     **BII:** $78°71$     Coordinates: [3]    unidentified

### General Data
D [kpc]: 10.4 kpc (M3) [2]      $N_H^{gal}$ [$10^{21}$ cm$^{-2}$]: 0.11
$L_{bol}$ [erg/s]: $1.2 \times 10^{34}$–$1.2 \times 10^{36}$ [3]     Orbital Period:
Mass of central object [$M_\odot$]:     Mass of companion [$M_\odot$]:
    Spectral type:

### X-ray Data
T [eV]: 20-45 (bb)     $N_H^{fit}$ [$10^{21}$ cm$^{-2}$]:
Orbital Modulation:
Variability: transient [3]

### Optical/UV Data
Finding Chart:     $m_B$ [mag]:     $m_V$ [mag]:
Orbital Modulation:
Opt. Spectrum:     UV Spectrum:
Opt. Variability:
UV Variability:
Nebula [erg/s]:     Wind mass loss:

### References
[1] Hertz P., Grindlay J.E., 1983, ApJ 275, 105: Einstein data, position, flux
[2] Webbink R.F., 1985, in IAU Symp. 123, Dynamics of Star Clusters, ed. J. Goodman & P. Hut (Dordrecht:Reidel), p. 541: distance to M3
[3] Hertz P., Grindlay J.E., Bailyn C.D., 1993, ApJ 410, L87: X-ray transient discovery in ROSAT HRI data, position, flux
[4] Fender R.P., Bell Burnell S.J., 1996, in Röntgenstrahlung from the Universe, Eds. H.U. Zimmermann et al., MPE Report 263, p. 135: infrared imaging (JHK)

## AG Dra                                            BD+67 922

**R.A.:** $16^h01^m40\overset{s}{.}9$     **LII:** $100\overset{\circ}{.}29$     Discovery: [10],[11]     **Sy**
**Dec.:** $+66°48'10''$     **BII:** $40\overset{\circ}{.}97$     Coordinates: [2]

### General Data
D [kpc]: 0.7–4 kpc     $N_H^{gal}$ [$10^{21}$ cm$^{-2}$]: 0.315
$L_{bol}$ [erg/s]: $1.4 \times 10^{36}$ (D/1 kpc)$^2$ [11]     Orbital Period: 554 d [5]
Mass of central object [$M_\odot$]:     Mass of companion [$M_\odot$]:
    Spectral type: K3III–K0Ib [3],[8]

### X-ray Data
T [eV]: 10–15 (bb) [10],[11]     $N_H^{fit}$ [$10^{21}$ cm$^{-2}$]: 0.4 [11]
Orbital Modulation: not clear [10],[11]
Variability: major drop during optical outbursts [10],[11]

### Optical/UV Data
Finding Chart: [1]     $m_B$ [mag]: 8–11.2     $m_V$ [mag]: 8–9.8
Orbital Modulation: 0.5 mag in U [5]
Opt. Spectrum: [3]     UV Spectrum: [6],[7]
Opt. Variability: series of outbursts, possibly 15 yr period [4]
UV Variability: factor 2–5 during outburst [7],[10]
Nebula [erg/s]:     Wind mass loss: 100 km/s [7]

## RR Tel                                                                 HV 3181

R.A.: $20^h04^m18\overset{s}{.}5$   LII: $342°\!.16$        Discovery: [9]           **Sy**
Dec.: $-55°43'34''$             BII: $-32°\!.24$        Coordinates: [3]

### General Data
D [kpc]: 2.6 kpc [8]                          $N_H^{gal}$ [$10^{21}$ cm$^{-2}$]: 0.439
$L_{bol}$ [erg/s]: $1.3\times10^{37}$ [9]     Orbital Period: 385–388 d [2],[4]
Mass of central object [$M_\odot$]: $>0.9$ [9]  Mass of companion [$M_\odot$]:
                                              Spectral type: M-giant [7]

### X-ray Data
T [eV]: 12 (wd) [9]                           $N_H^{fit}$ [$10^{21}$ cm$^{-2}$]: 0.17
Orbital Modulation:
Variability:

### Optical/UV Data
Finding Chart: [3]              $m_B$ [mag]:7–14              $m_V$ [mag]:7–14
Orbital Modulation: 2.5 mag amplitude [2]
Opt. Spectrum: [5]                            UV Spectrum: [6]
Opt. Variability: outburst 1944/45 and decline [2],[3]
UV Variability: marginal
Nebula [erg/s]:                               Wind mass loss:

### References
[1] Pickering E.C., 1908, Harvard Coll. Obs. Circ. No. 143: discovery of variability by Mrs. Fleming
[2] Payne C.H., 1928, Bull. Harv. Coll. Obs. No. 861: long-term optical variations
[3] Mayall M.W., 1949, Harv. Coll. Obs. Bull. No. 919, p. 15: finding chart
[4] Gaposchkin S., 1945, Ann. Harv. Coll. Obs. No. 2: orbital period, long-term lightcurve
[5] Thackeray A.D., 1977, Mem. Roy. Astr. Soc. 83, 1: optical spectra 1951–1973, nebular evolution
[6] Penston M.V., Benvenuti P., Cassatella A., Heck A., Selvelli P., Macchetto F., Ponz D., Jordan C., Cramer N., Rufener F., Manfroid J., MNRAS 202, 833: IUE spectra and line identification, temperature and density
[7] Feast M.W., Whitelock P.A., Catchpole R.M., Robertson B.S.C., Carter B.S., 1983, MNRAS 202,951: JHKL photometry 1972–1981, Mira identification
[8] Whitelock P.A., 1988, in The Symbiotic Phenomenon, IAU Coll. 103, ed. J. Mikolajewska, M. Friedjung, S.J. Kenyon, R. Viotti, Kluwer, p. 47: distance
[9] Jordan S., Mürset U., Werner K., 1994, A&A 283, 475: X-ray discovery, atmosphere modelling of X-ray, UV and optical data

# V1974 Cyg  Nova Cyg 1992

R.A.: $20^h30^m31^s2$  LII: $345°92$  Discovery: [11]  N
Dec.: $-52°37'53''$  BII: $-36°06$  Coordinates: [1]

### General Data
D [kpc]: 1.8–3.2 [12]  $N_H^{gal}$ [$10^{21}$ cm$^{-2}$]: 0.325
$L_{bol}$ [erg/s]: $2\times 10^{38}$ [6]  Orbital Period: 1.95 h [9]
Mass of central object [$M_\odot$]: 0.75–1.1 [12],[13]  Mass of companion [$M_\odot$]:
    Spectral type:

### X-ray Data
T [eV]:  $N_H^{fit}$ [$10^{21}$ cm$^{-2}$]:
Orbital Modulation:
Variability: full outburst lightcurve [11]

### Optical/UV Data
Finding Chart: [2]  $m_B$ [mag]:  $m_V$ [mag]: 4.5–21
Orbital Modulation: 0.2 mag [9]
Opt. Spectrum: [4],[10]  UV Spectrum: [5]
Opt. Variability:
UV Variability: [5],[7]
Nebula [erg/s]:  Wind mass loss:

### References
[1] Collins P., Skiff B.A., 1992, IAUC 5454: optical discovery, coordinate
[2] A.A.V.S.O. Circ. No. 256, 1992: finding chart
[3] Shore S.N., Sonneborn G., Starrfield S., Gonzalez-Riestra R., Ake T.B., 1993, AJ 106, 2408: nebular evolution from HST + IUE spectra, HI limit from Ly$\alpha$ P Cyg profile, ejected mass
[4] Barger A.J., Gallagher J.S., Bjorkman K.S., Johansen K.A., Nordsieck K.H., 1993, ApJ 419: optical spectra during decline
[5] Shore S.N., Sonneborn G., Starrfield S., Gonzalez-Riestra R., Ake T.B., 1993, AJ 106, 2408: UV spectral evolution
[6] Shore S.N., Sonneborn G., Starrfield S., Gonzalez-Riestra R., Polidan R.S., 1994, ApJ 421, 344: IUE observations, UV and V lightcurve during first year, bolometric evolution
[7] Taylor M., Bless R.C., Ögelman H., Elliot J.L., Gallagher J.S., Nelson M.J., Percival J.W., Robinson E.L., van Citters G.W., 1994, ApJ 424, L45: UV photometry
[8] Bjorkman K.S., Johansen K.A., Nordsieck K.H., Gallagher J.S., Barger A.J., 1994, ApJ 425, 247: spectropolarimetry
[9] DeYoung J.A., Schmidt R.E., 1994, ApJ 431, L47: orbital period
[10] Rafanelli P, Rosino L., Radovich M., 1995, A&A 294, 488: optical spectra during outburst
[11] Krautter J., Ögelman H., Starrfield S., Wichmann R., Pfeffermann E., 1996, ApJ 456, 788: X-ray rise and decline, X-ray spectrum
[12] Paresce F., Livio M., Hack W., Korista K., 1996, A&A 299, 823: nebula resolved with HST, UV spectra, abundances, distance WD mass
[13] Krautter J., 1996 (priv. comm.)